\documentclass{svproc}
\usepackage{url}
\usepackage{graphicx,xcolor}
\usepackage{amsmath}
\usepackage{amsfonts}
\usepackage{amssymb}
\usepackage{amstext}
\usepackage{caption}
\usepackage[symbol]{footmisc}

\begin{document}
\mainmatter              
\title{Can we measure $\theta_{23}$ octant in 3+1 scheme?}
\titlerunning{Octant of $\theta_{23}$ }  
%
\author{Sanjib Kumar Agarwalla\inst{1,2,\ast} \and Sabya Sachi Chatterjee\inst{1,2,\dagger,}$\footnote[4]{Speaker, corresponding author.}$ 
\and Antonio Palazzo\inst{3,4,\ddagger}}
\authorrunning{Agarwalla et al.} 
%
\tocauthor{Sanjib Kumar Agarwalla, Sabya Sachi Chatterjee, and
 Antonio Palazzo}
\institute{Institute of Physics, Sachivalaya Marg, Sainik School Post, Bhubaneswar 751005, India\\
\and Homi Bhabha National Institute, Training School Complex, Anushakti Nagar, Mumbai 400085, India\\
\and Dipartimento Interateneo di Fisica ``Michelangelo Merlin,'' Via Amendola 173, 70126 Bari, Italy\\
\and Istituto Nazionale di Fisica Nucleare, Sezione di Bari, Via Orabona 4, 70126 Bari, Italy\\
\email{$^\ast$sanjib@iopb.res.in, $^\dagger$sabya@iopb.res.in, $^\ddagger$palazzo@ba.infn.it}
 }

\maketitle              

\begin{abstract}
Current 3$\nu$ global fits predict two degenerate solutions for $\theta_{23}$: one lies in lower octant ($\theta_{23} <\pi/4$), and the other belongs to higher octant ($\theta_{23} >\pi/4$). Here, we study how the measurement of $\theta_{23}$ octant would be affected in the upcoming Deep Underground Neutrino Experiment (DUNE) if there exist a light eV-scale sterile neutrino. We show that in 3+1 scheme, a new interference term in $\nu_\mu \to \nu_e$ oscillation probability  can spoil the chances of measuring $\theta_{23}$ octant completely.


\keywords{Octant of $\theta_{23}$, sterile neutrino, Long-Baseline experiments}
\end{abstract}

{\bf {\em Introduction:}}
The resolution of octant\footnote{According to the present 3$\nu$ best-fit \cite{Capozzi:2016rtj}, $\theta_{23}$ can have two solutions: one $<$ $\pi/4$, labelled as lower octant (LO), and other $>$ $\pi/4$, known as higher octant (HO).} \cite{Fogli:1996pv} of $\theta_{23}$ is one of the fundamental problems in neutrino oscillation. Long-baseline (LBL) experiments \cite{Agarwalla:2013ju} can resolve this octant ambiguity of $\theta_{23}$ with the help of $\nu_{\mu}\rightarrow \nu_e$ appearance channel, and the vital information coming from $\nu_{\mu}\rightarrow \nu_{\mu}$ disappearance channel also play an important role. Interestingly, at present, there are short-baseline anomalies which hint towards the existence of light eV-scale sterile neutrino \cite{Abazajian:2012ys}. Here, we expound in detail the capability of proposed LBL experiment DUNE to measure $\theta_{23}$ octant considering one light eV-scale sterile neutrino along with three active neutrinos.


{\bf {\em Theoretical framework:}} In the 3+1 scheme, a new a mass eigenstate $\nu_4$ appears on top of 3$\nu$ framework whose mixing is parametrized as
\begin{equation}
\label{eq:U}
U =   \tilde R_{34}  R_{24} \tilde R_{14} R_{23} \tilde R_{13} R_{12}\,, 
\end{equation} 
where $R_{ij}$ ($\tilde R_{ij}$) is a real (complex) rotation in the ($i,j$) plane. The details of the parametrization of U can be seen in \cite{Klop:2014ima}. 

In~\cite{Klop:2014ima}, it was shown that the 4-flavor appearance probability can be 
approximately expressed as the sum of three terms $P^{4\nu}_{\mu e}  \simeq  P_{\rm{0}} + P_{\rm {1}}+   P_{\rm {2}}$,\,
which in vacuum appears as
\begin{eqnarray}
\label{eq:Pme_atm}
 & P_{\rm {0}} &\!\! \simeq\,  4 s_{23}^2 s^2_{13}  \sin^2{\Delta}\,,\\
\label{eq:Pme_int_1}
 & P_{\rm {1}} &\!\!  \simeq\,   8 s_{13} s_{12} c_{12} s_{23} c_{23} (\alpha \Delta)\sin \Delta \cos({\Delta \pm \delta_{13}})\,,\\
 \label{eq:Pme_int_2}
 & P_{\rm {2}} &\!\!  \simeq\,   4 s_{14} s_{24} s_{13} s_{23} \sin\Delta \sin (\Delta \pm \delta_{13} \mp \delta_{14})\,.
\end{eqnarray}
where $\Delta \equiv  \Delta m^2_{31}L/4E$ 
and
$\alpha \equiv \Delta m^2_{21}/ \Delta m^2_{31}$.
In the double sign, the upper (lower) sign corresponds to neutrinos (antineutrinos).
The new interference term  $P_{\rm {2}}$ is governed by the interference between the atmospheric frequency and the large frequency related to the new mass eigenstate~\cite{Klop:2014ima} which gets averaged out by the finite energy resolution of the detector. Recent global fits \cite{Capozzi:2016rtj,Giunti:2013aea,Kopp:2013vaa} suggests $s_{13}\sim s_{14} \sim s_{24} \sim 0.15\, (\sim \epsilon)$ and $\alpha \,=\, 0.03\, (\sim \epsilon^2)$ implying $P_{\rm {0}} \sim \epsilon^2,\,\;P_{\rm {1}} \sim \epsilon^3,\,\;P_{\rm {2}} \sim \epsilon^3$.
An experiment can measure the octant of $\theta_{23}$ even in the presence of unknown CP-phases, if there is a difference between the probabilities corresponding to the different octants, i.e.
\begin{eqnarray}
\label{eq:DPme}
\Delta P \equiv P^{\mathrm {HO}}_{\mu e} (\delta_{13}^{\mathrm {HO}}, \delta_{14}^{\mathrm {HO}}) -
                           P^{\mathrm {LO}}_{\mu e} (\delta_{13}^{\mathrm {LO}}, \delta_{14}^{\mathrm {LO}})\ne 0\,,
\end{eqnarray}

where one of the two octants should be considered to generate data and the other octant should be used to simulate the theoretical model. From the expression of $P^{4\nu}_{\mu e}$,
$\Delta P$ can be written as,\,\;\; $\Delta P =   \Delta P_{\rm{0}} + \Delta P_{\rm {1}} +   \Delta P_{\rm {2}}$.

$\Delta P_{\rm{0}}$ is positive-definite. $\Delta P_{\rm {1}}$ and $\Delta P_{\rm {2}}$ depends on the CP-phases and can be both positive or negative. These terms can be expressed as 
\begin{eqnarray}
\label{eq:DP_0}
\Delta P_{\rm {0}}& \simeq & 8 \eta s_{13}^2 \sin^2\Delta\,.\\
\Delta P_{\rm {1}}& =& 4 s_{13} s_{12} c_{12} (\alpha \Delta)\sin \Delta \big[ \cos(\Delta \pm \phi^{\mathrm {HO}}) - \cos(\Delta \pm \phi^{\mathrm {LO}})\big] \,,\\
\label{eq:DP_2}
\Delta P_{\rm {2}} &= & 2 \sqrt{2} s_{14} s_{24} s_{13} \sin\Delta \big[ \sin(\Delta \pm \psi^{\mathrm {HO}}) - \sin(\Delta \pm \psi^{\mathrm {LO}})\big]\,,
\end{eqnarray}

In above, $\phi\, = \, \delta_{13}$ and $\psi \,=\,  \delta_{13} -  \delta_{14}$ with the appropriate superscripts LO or HO. The term $\eta$ is a positive definite angle and dictates the deviation from maximal mixing as $\theta_{23} = \frac{\pi}{4}\pm \eta$, where +(-) corresponds to HO (LO). If we need to measure the  octant of $\theta_{23}$, the contribution coming from $\Delta P_{\rm {0}}$ must not get cancelled completely in cases where the sum of $\Delta P_{\rm {1}}$ and $\Delta P_{\rm {2}}$ gives a negative contribution.

\begin{figure}[h!]
\vspace*{-0.0cm}
\hspace*{-0.3cm}
\center\includegraphics[width=6. cm]{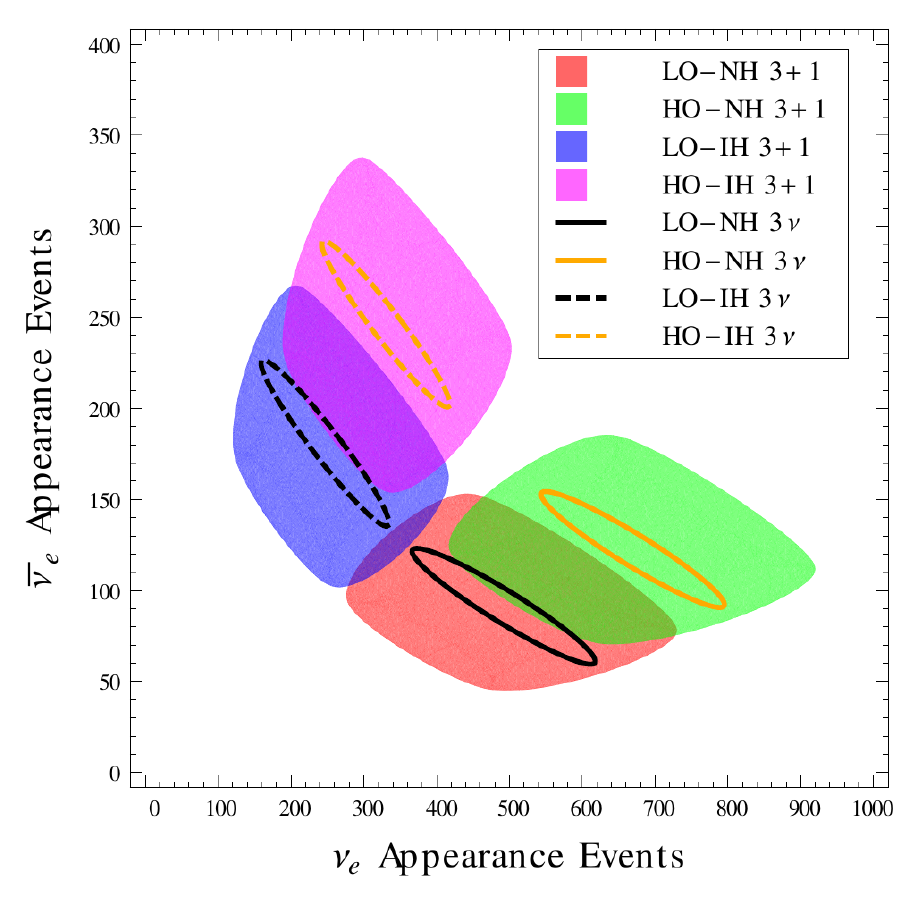}
\vspace*{-0.0cm}
\caption{Bi-event plot for DUNE. The ellipses (colored blobs) corresond to the 3$\nu$ (3+1 scheme).
$\sin^2\theta_{23} = 0.42$\,(0.58) is taken as benchmark value for LO (HO). In cases of ellipses (colored blobs), the running parameter(s) is $\delta_{13}$ ($\delta_{13}$ \& $\delta_{14}$). In the 3+1 case, $\theta_{34}$ = $0^0$ has been assumed. This figure has been taken from \cite{Agarwalla:2016xlg}.
\label{fig:bievents}}
\end{figure}  

{\bf {\em Results and discussion:}}
Simulations for DUNE have been performed considering a total 248 kt.MW.yr of exposure, divided equally between $\nu$ and $\bar{\nu}$ mode. In fig.\ref{fig:bievents}, we show 
the bi-event plot. The ellipses (colored blobs) correspond to the 3$\nu$ (3+1 scheme) where, $\sin^2 \theta_{23} = 0.42\,(0.58)$ has been assumed as a benchmark value for
the LO (HO). Since mass hierarchy can be measured relatively easily\footnote{In fig.\ref{fig:bievents}, we notice an small overlap between normal hierarchy (NH) and inverted hierarchy (IH) blobs which can be eliminated using the spectral information available in DUNE (see~\cite{Agarwalla:2016xxa}).
} in DUNE, we can only concentrate on one of the two hierarchies, say normal hierachy. 
While going from 3$\nu$ to 3+1 scheme, the ellipses becomes blobs because of the convolution of different combinations of $\delta_{13}$ \& $\delta_{14}$ (see~\cite{Agarwalla:2016xxa,Agarwalla:2016mrc}). In this figure, we see a substantial overlap between LO and HO blobs due to the presence of the term $\Delta P_2$, 
which depends on the new CP-phase $\delta_{14}$. For detailed discussion, see \cite{Agarwalla:2016xlg}.

%
 
 \begin{figure}[h!]
\vspace*{-0.5cm}
\hspace*{-0.5cm}
\center\includegraphics[width=11.2 cm]{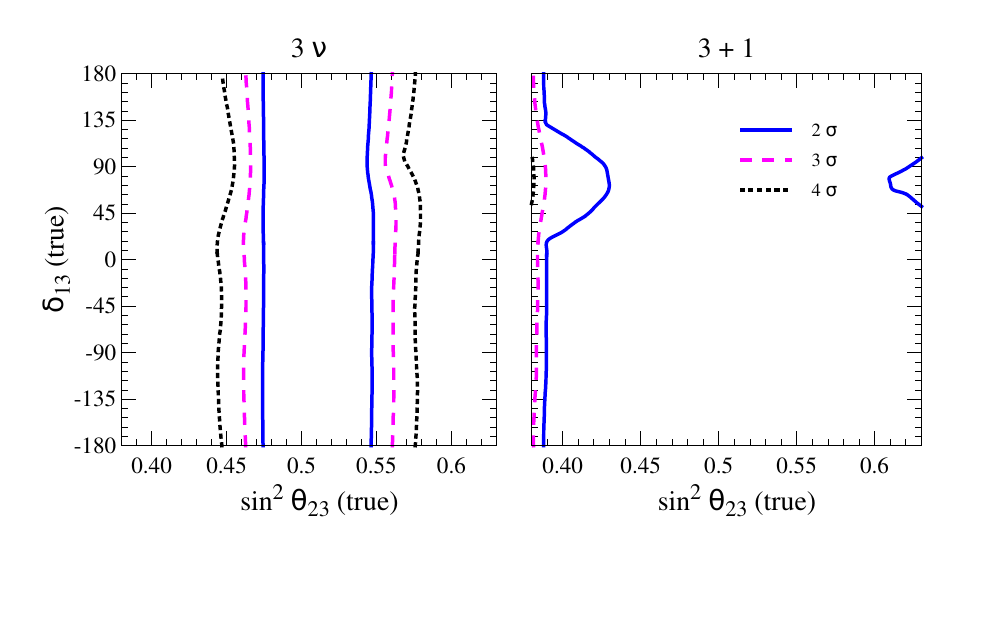}
\vspace*{-1.3cm}
\caption{Discovery potential for excluding the wrong octant in [$\sin^2 \theta_{23}, \delta_{13}$] (true) plane
assuming NH as true choice.
The left (right) panel corresponds to the 3$\nu$ (3+1) case.
In 3-flavor scenario, we marginalize over ($\theta_{23}, \delta_{13}$) (test). In 3+1 case, 
in addition, we marginalize over $\delta_{14}$ (true) and $\delta_{14}$ (test) fixing
$\theta_{14} = \theta_{24} = 9^0$ and $\theta_{34} = 0$. This figure has been taken from \cite{Agarwalla:2016xlg}.
\label{fig:2pan_octant_true}}
\end{figure}  

 Fig.\ref{fig:2pan_octant_true} depicts the discovery reach of $\theta_{23}$ octant in $\left[\sin^2\theta_{23}, \delta_{13}\right]$(true) plane assuming NH as true choice.
Left (right) panel shows the results for 3$\nu$ (3+1) scheme. In 3$\nu$ case, a minimum 2$\sigma$ sensitivity can be achieved if $\sin^2\theta_{23}(\rm true)\,\lesssim\, 0.47$ and $\sin^2\theta_{23}(\rm true)\,\gtrsim\, 0.55$ irrespective of the choice of $\delta_{13}$(true). But, in 3+1 case we hardly have any octant sensitivity in the entire $\left[\sin^2\theta_{23}, \delta_{13}\right]$(true) plane.


{\bf {\em Conclusions:}}
In this work, we have studied the impact of a light eV-scale sterile neutrino in measuring the octant of $\theta_{23}$ at DUNE. The sensitivity towards $\theta_{23}$ octant can be completely lost if there is active- sterile oscillations.

{\bf {Acknowledgments:}}
S.S.C. would like to thank the organizers of XXII DAE-BRNS HEP Symposium 2016 for giving an opportunity to present this work. S.K.A. is supported by the DST/INSPIRE Research Grant [IFA-PH-12],
Department of Science \& Technology, India.
A.P. is supported 
by the grant ``Future In Research''  {\it Beyond three neutrino families}, 
Fondo di Sviluppo e Coesione 2007-2013, 
APQ Ricerca Regione Puglia, Italy, ÒProgramma regionale a sostegno della
specializzazione intelligente e della sostenibilit\`a sociale ed ambientale. 
A.P. acknowledges partial support by the research project 
{\em TAsP} funded by the Instituto Nazionale di Fisica Nucleare (INFN). 

%
%

\end{document}